\documentclass[twocolumn,floats,prb,aps,showpacs]{revtex4}

\usepackage[dvips]{graphicx}
\usepackage{amssymb}
\usepackage{euscript}

\newcommand{\icm}{\ensuremath{~\textrm{cm}^{-1}}}	
\newcommand{\etal}{{\it et al.}}					
\newcommand{\hand}[1]{\ensuremath{{\EuScript #1}}}	

\begin{document}


\title{Photoinduced time-resolved electrodynamics of superconducting metals and alloys}

\author{R.P.S.M. Lobo}
\email{lobo@espci.fr}
\affiliation{Laboratoire de Physique du Solide, CNRS UPR 5, Ecole Sup\'erieure de Physique et Chimie Industrielles de la Ville de Paris, 75231 Paris cedex 5, France}

\author{J.D. LaVeigne}
\author{D.H. Reitze}
\author{D.B. Tanner}
\affiliation{Department of Physics, University of Florida, Gainesville, FL 32611-8440}

\author{Z.H. Barber}
\affiliation{Department of Materials Science and Metallurgy, University of Cambridge, Pembroke Street, Cambridge CB2 3QZ, United Kingdom}

\author{E. Jacques}
\author{P. Bosland}
\affiliation{DAPNIA, Commissariat \`a l'Energie Atomique - Saclay, 91191 Gif-sur-Yvette, France}

\author{M.J. Burns}
\affiliation{Jet Propulsion Laboratory, California Institute of Technology, Pasadena, CA 91109} 

\author{G.L. Carr}
\affiliation{National Synchrotron Light Source, Brookhaven National Laboratory, Upton, NY 11973}

\date{\today}

\begin{abstract}
The photoexcited state in superconducting metals and alloys was studied {\it via\/} pump-probe spectroscopy. A pulsed Ti:sapphire laser was used to create the non-equilibrium state and the far-infrared pulses of a synchrotron storage ring, to which the laser is synchronized, measured the changes in the material optical properties. Both the time- and frequency- dependent photoinduced spectra of Pb, Nb, NbN, Nb$_{0.5}$Ti$_{0.5}$N, and Pb$_{0.75}$Bi$_{0.25}$ superconducting thin films were measured in the low-fluence regime. The time dependent data establish the regions where the relaxation rate is dominated either by the phonon escape time (phonon bottleneck effect) or by the intrinsic quasiparticle recombination time. The photoinduced spectra measure directly the reduction of the superconducting gap due to an excess number of quasiparticles created by the short laser pulses. This gap shift allows us to establish the temperature range over which the low fluence approximation is valid. 
\end{abstract}

\pacs{74.40.+k, 74.70.Ad, 74.25.Gz, 78.47.+p}

\maketitle


\section{INTRODUCTION}
\label{intro}

When a superconductor is subjected to a pulse of above-bandgap electromagnetic radiation, excess quasiparticles (QPs) are created. The number of excess QPs created, and the way in which they recombine back into Cooper pairs, are processes of fundamental importance. Burstein and co-workers\cite{Burstein1961} made the first attempt to estimate this recombination rate. Assuming a radiative process (emission of a photon) they calculated a recombination time around 0.4 s. Ginsberg's\cite{Ginsberg1962} tunneling data in a Pb film, into which excess QPs had been injected, showed recombination times many orders of magnitude smaller ($10^{-7}$ s) giving indications that the process was dominated by emission of phonons rather than photons. Subsequent calculations,\cite{Schrieffer1962,Rothwarf1963,Rothwarf1967,Gray1969} confirmed the faster recombination rate when phonon emission is considered. However, Rothwarf and Taylor\cite{Rothwarf1967} showed that the measured (or effective) lifetime was longer than the ``bare'' or intrinsic recombination time due to phonon-trapping effects.

Many groups have studied the problem of nonequilibrium superconductivity by injection of unpaired quasiparticles in tunnel junctions.\cite{Miller1967,Levine1968,Gray1969b,Gray1971,Tinkham1972} In these steady-state measurements, the excited state is detected through an average response that deviates from equilibrium and the effective pair relaxation time is inferred, assuming that the process has a simple exponential decay.

As an alternative, the non-equilibrium state can be generated by breaking Cooper pairs. Testardi\cite{Testardi1971} showed that light of enough intensity could destroy the superconductivity in a process not attributable to lattice heating. In this process, photons with energy larger than $2 \Delta$ ($\Delta$ being the superconducting gap) break pairs. Parker and Williams\cite{Parker1972} used this phenomenon to measure pair recombination rates in illuminated tunnel junctions.

Owen and Scalapino\cite{Owen1972} described the non-equilibrium state by an effective chemical potential ($\mu^\star$) and showed that the excited state is characterized by a shift in the superconducting gap. Sai-Halasz \etal\cite{Sai1974} used a pulsed laser to create the excess number of QPs and analyzed the changes in the microwave reflectivity. Their data support the theory of Owen and Scalapino for the low-fluence regime, where the number of QPs created by the photon absorption is much smaller than the number of unpaired QPs produced by finite sample temperature. The gap shift in the photoexcited state was also detected by Hu \etal\cite{Hu1974} in tunnel junctions; this experiment, which had a time resolution of 30 ns, was the first direct measurement of the relaxation process. Parker then proposed that the photoexcited state could be described by BCS theory with an effective temperature ($T^\star$) based on the phonons with energy greater than $2 \Delta$.\cite{Parker1975} For small perturbations both $T^\star$ and $\mu^\star$ models are indistinguishable.

The major drawback of most previous work is the indirect manner by which the relaxation rates were measured. Johnson\cite{Johnson1991} was the first to use fast lasers and electronics to look at voltage transients with time resolution in the range of 100~ps. He directly measured the effective relaxation times in Nb films and found a relaxation composed of a double exponential decay, suggested to be intrinsic. Federici \etal\cite{Federici1992} used coherent THz techniques to look at the pair-breaking dynamics, finding that pairs are broken in picoseconds.

More recently, all optical ultra-fast pump-probe techniques were used to study the non-equilibrium dynamics of high-$T_c$ superconductors.\cite{Han1990,Reitze1992,Feenstra1997,Kabanov1999,Averitt2001,Kaindl2002,Segre2002,Gedik2004} In these materials, relaxation times were three orders of magnitude faster than in metallic superconductors, but no direct gap shift was measured. Although the gap symmetry of the high-$T_c$ materials plays a major role in the relaxation times,\cite{Howell2004} qualitatively the process is similar to the one observed in BCS superconductors.\cite{Carbotte2004} All optical pump-probe spectroscopy also proved to be very useful in the understanding of the non-equilibrium dynamics of the double gapped
MgB$_2$.\cite{Xu2003,Demsar2003,Lobo2004}

In this paper we extend our far-infrared pump-probe spectroscopy work done on Pb thin films\cite{Carr2000} to Nb, NbN, Nb$_{0.5}$Ti$_{0.5}$N and Pb$_{0.75}$Bi$_{0.25}$ and we develop several issues linked to the recombination process. This paper is divided as follows: in Sec.~\ref{sec_experim} we describe our samples, the experimental setup
and the characterization methods used. In Sec.~\ref{sec_eqfir} we discuss the equilibrium far-infrared spectra of our samples. Section \ref{sec_excessQP} deals with the non-equilibrium excess QP state while our conclusions are presented in Sec.~\ref{sec_conc}. In the Appendix we discuss the consequences of sample heating by the laser and methods for minimizing this effect.

\section{EXPERIMENTAL}
\label{sec_experim}

\subsection{Samples}
\label{subsec_samples}

Our samples were thin metallic films of Pb, Pb$_{0.75}$Bi$_{0.25}$, Nb, NbN, and Nb$_{0.5}$Ti$_{0.5}$N. All these metals were deposited on 0.5 mm thick c-axis sapphire, except for the NbN, which was deposited on 0.5 mm thick (100) MgO crystal. The film thickness were not directly determined. As it will be discussed in Sec.\ \ref{sec_eqfir}, the important quantity is the sheet resistance $R_\Box$ which can be calculated from the normal state transmission. From $R_\Box$ we can estimate roughly the thicknesses of our films, obtaining values ranging from 50 to 200 \AA. Several films of each composition were grown.

Generally, the far-infrared transmission decreases exponentially with the film thickness, therefore decreasing the signal to noise ratio. However, $T_c$ is usually higher for thicker films. In this paper we show in detail data for the films with the highest 
$T_c$'s consistent with a normal-state far-infrared transmission between 10\% and 30\%.

Pb and Pb$_{0.25}$Bi$_{0.75}$ films were evaporated {\it in situ} at 80~K and the samples were kept below 100~K throughout the measurements to avoid annealing and normal state conductivity change. The transition temperatures were optically determined (see `Methods'
below) by the increase in the far-infrared transmission that begins at $T_c$. Twenty Pb
films were deposited and we measured $T_c$ between 7 and 7.2~K. The four Pb$_{0.25}$Bi$_{0.75}$ films showed the superconducting transition at 8.0~K. The $T_c$'s in these samples were independent of the deposition time and hence of the film thickness.

Nb films were sputtered and kept in vacuum to avoid oxidation and degradation of the superconducting properties. Two different samples showed (optically determined) $T_c$ of 5.9~K and 5.4~K. We show here the data for the latter. NbN was produced by reactive sputtering of Nb in N$_2$ atmosphere.\cite{Barber1993,Somal1996} The sample had $T_c = 13.5$~K. Nb$_{0.5}$Ti$_{0.5}$N films were sputtered in an N$_2$ atmosphere. For thick films (in
the $\mu$m range) this technique produces samples with $T_c \approx 14.5$~K. Our films were much thinner (100--200 \AA) and showed suppressed $T_c$'s, below 11~K.

\subsection{Apparatus}
\label{subsec_Apparatus}

In our measurements we utilized the far-infrared photons from beamline U12IR in the VUV synchrotron storage ring of the National Synchrotron Light Source at BNL.\cite{Lobo1999}

Spectrally-resolved measurements were recorded from 5 to 60\icm\ with a lamellar grating interferometer\cite{Henry1979} for the Pb, Pb$_{0.25}$Bi$_{0.75}$ and Nb$_{0.5}$Ti$_{0.5}$N films. A Bruker IFS~66v was utilized to measure Nb and NbN samples from 5 to 100\icm. Most of the data were obtained with a 1.3~K bolometer detector although a few measurements were taken with a 4.2~K bolometer. Black polyethylene and fluorogold filters gave an upper frequency limit of 100\icm. The equilibrium transmittance spectra of NbN was measured at ESPCI in a Bruker IFS~66v using a Hg-arc source and a home-made cryostat.

Low-temperature transmittance data for the Pb and Pb$_{0.25}$Bi$_{0.75}$ films were measured using a ARS He-Tran cold finger cryostat. The samples (10 $\times$ 10 mm$^2$) were solidly clamped to the cold finger using indium gaskets. A 5 mm diameter circular aperture was used, and the remainder of each sample remained in contact with the cold surface. The NbN and Nb$_{0.5}$Ti$_{0.5}$N films were measured in an Oxford Optistat cryostat, with the samples immersed in cold gas or liquid helium. The Nb films were measured in both cryostats. In some experiments, we measured the temperature-dependence of the overall far-infrared transmission (without frequency discrimination by a spectrometer) through the sample. In this case we modulated the beam at around 100 Hz with an optical chopper and used a lock-in amplifier.

To create the non-equilibrium photoinduced state, a mode-locked Ti:Sapphire laser was synchronized to the rf cavity of the VUV synchrotron storage ring. The laser produces a maximum of 1~W average power distributed in pulses with a repetition rate of 52~MHz, the same repetition rate as the synchrotron pulses. The laser pulses are approximatelly 2--3~ps long at a photon energy of 1.5 eV. The laser is used to produce excess quasiparticles in the superconductor and the far-infrared synchrotron pulses probe the changes in the sample transmission caused by these broken pairs. The time resolution is determined by the synchrotron pulse width, and is $\sim 200$ ps. The interpulse separation limits the window for relaxation to 19 ns. More details on the facility can be found in Ref.~\onlinecite{Lobo2002}. Typical laser powers used in our experiments were in the range of 10 to 100 mW (0.2 to 2 nJ per pulse) evenly distributed over the 5 mm clear aperture on the samples. Such low power is required to minimize heating and to keep the material in the low-fluence regime.

\subsection{Methods}
\label{subsec_methods}

\begin{figure}
  \begin{center}
    \includegraphics[width=8cm]{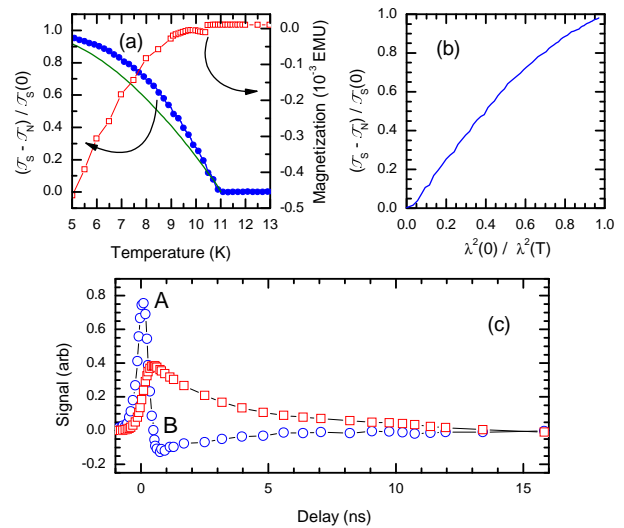}
  \end{center}
\caption{(Color online) (a) The left axis shows the increase in the far-infrared transmission due to the superconducting transition in one of our Nb$_{0.5}$Ti$_{0.5}$N samples. The solid line is a BCS calculation for the temperature evolution of the penetration depth $\lambda^2(0) / \lambda^2(T)$, a quantity which is proportional to the number of pairs in the system. On the right axis we plot the magnetization of the same sample. Note the suppressed zero. (b) Overall far-infrared transmission vs.  calculated $\lambda^2(0) / \lambda^2(T)$. (c) Differential transmission measurement ($\circ$) and its integral ({$\scriptscriptstyle\Box$}) for a Nb$_{0.5}$Ti$_{0.5}$N film at 2~K. Point `A' in the differential curve corresponds to the maximum up slope in the integrated curve and point `B' to its maximum down slope.}
\label{fig_TcNqpZPD}
\end{figure}
Two types of transmission were employed: (i) we recorded the overall far-infrared light transmitted through the sample without energy discrimination and (ii) we measured frequency-resolved spectra between $\sim 5$ and 100\icm (60\icm\ for measurements in the lamellar interferometer). The overall transmitted light is obtained simply by placing the sample in the optical beam between the source and the detector. This overall transmission is the integrated sample transmission below 100\icm, weighted by the emission spectrum of the synchrotron and by the transmission of the filters, the cryostat windows, and the low-pass filter in the bolometer cryostat. So long as the cut-off frequency is higher than about $3 \Delta$ (which is our case), the transmission of a superconductor increases in the superconducting state. This property can be used to measure the sample $T_c$. Figure \ref{fig_TcNqpZPD}(a) compares the increase in the overall far-infrared transmission to the magnetization measured in a Quantum Design MPMS squid magnetometer for one of our samples. The optically-determined $T_c$ is about 0.5~K higher than the value obtained from the magnetic measurement. This discrepancy is attributed to the fact that the far-infrared transmission senses resistive transitions, and percolation of one superconducting path may happen before the entire sample becomes superconducting. As shown in Fig.\ \ref{fig_TcNqpZPD}(b) this overall transmission can also be easily related to the number of Cooper pairs in the system. The curve in Fig.\ \ref{fig_TcNqpZPD}(b) is not universal and must be determined for each sample measured. Note that it is approximately but not exactly linear.

The non-equilibrium state created by the laser pulses was, for the most part, studied using the overall transmitted light through the sample. These data allowed us to measure the variation of the transmission as a function of the pump and probe delay, yielding the temperature dependence of the amplitude of the photoinduced signal and its relaxation time. 

\begin{figure*}
  \begin{center}
    \includegraphics[width=16cm]{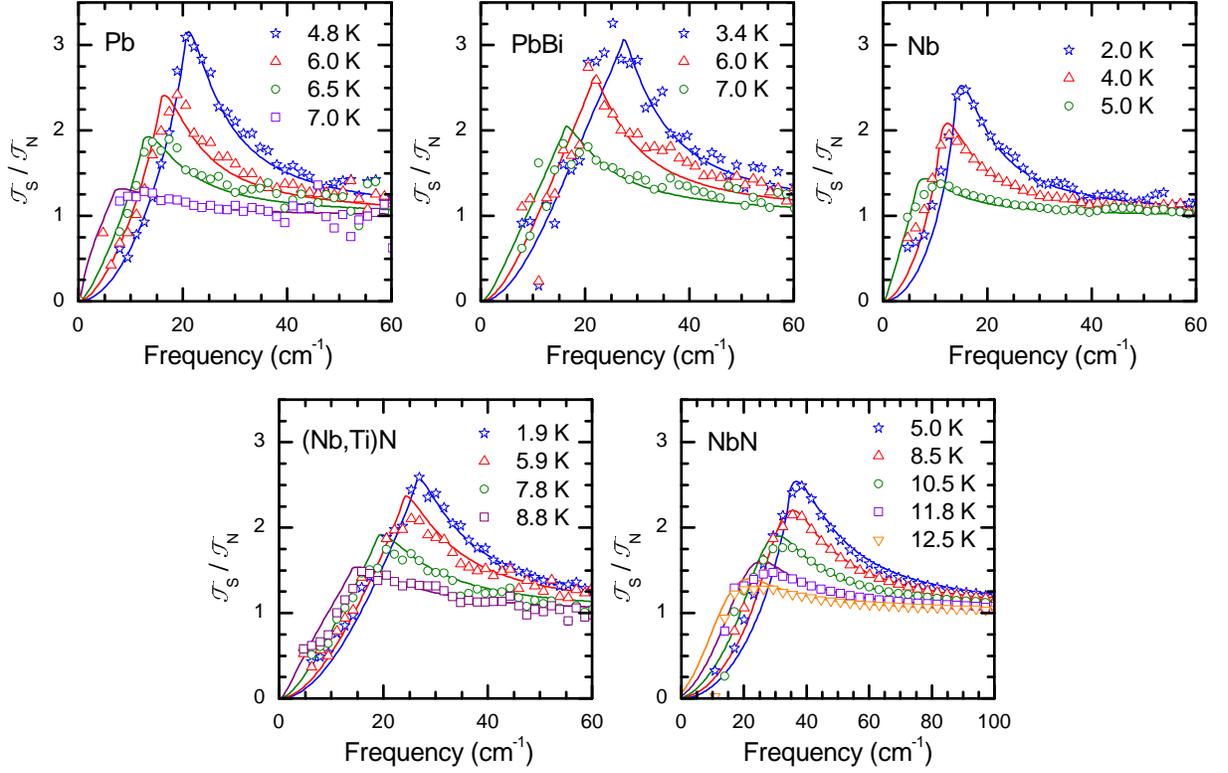}
  \end{center}
\caption{(Color online) Ratio of superconducting-state to normal-state far-infrared transmission for several temperatures. The symbols are the experimental data and the lines are strong-coupling BCS fits using the parameters given in Table \ref{tb_TsTn}. Note that the frequency scale for the NbN film is broader than for the other samples.}
\label{fig_TsTn}
\end{figure*}
\begin{table*}
  \begin{center}
    \begin{tabular}{lccccccccc}
      \hline
      \hline
      Material & $R_\Box (\Omega)$ & $T_c$ (K) & $2\Delta_0(\icm)$ &
            $2\Delta_0 / k_B T_c$ & $\delta S (\Omega^{-1} \textrm{cm}^{-2})$ &
           $\tau_R^0 / \tau_B^0$ & $I$ (nJ / pulse) & $N_{qp} / N_T$\\
      \hline
      Pb & 70 & 7.2 & 22.5 & 4.5 &  -0.5 & 2.53 & 0.5 & 0.03 (3.7 K)\\
      Pb$_{0.75}$Bi$_{0.25}$ & 110 & 8.0 & 28 & 4.9 &  -0.7 & 2.2 & 0.5 & 0.24 (3.0 K) \\
      Nb & 61 & 5.4 & 14 & 3.7 &  0 & 13.3 & 2 & 0.03 (3.5 K) \\
      Nb$_{0.5}$Ti$_{0.5}$N & 110 & 9.8 & 26.5 & 3.9 &  -0.2 & 6 & 2 & 5 (2.0 K) \\
      NbN & 70 & 13.5 & 35.0 & 3.7 &  0 & 1.53 &  2 & 14 (2.2 K) \\
      \hline
      \hline
    \end{tabular}
  \end{center}
\caption{Parameters for Mattis-Bardeen fits of the far infrared superconducting to normal transmission ratio $\hand{T}_s / \hand{T}_n$. In all cases we used the dirty limit ($1 / \tau \rightarrow \infty$) approximation. The time ratio obtained from the non-equilibrium fits, the laser fluences (filling the 5 mm sample aperture) utilized and the excess QP fraction $N_{qp} / N_T$ are also shown. For the latter the number in parenthesis is the  lowest sample temperature reached.}
\label{tb_TsTn}
\end{table*}

We use a differential technique to determine the time dependence of the pair recombination. After setting the pump-probe delay ($t$) a square wave signal around 100 Hz is fed to the laser electronics making the laser pulses hop between two time locations. The time separation ($\delta t$) between these two locations can be set to any value up to 600 ps. The modulation signal then serves as the reference for lock-in detection of the far-infrared signal. The signal is the difference in the transmission of two excited states occurring at pump to probe delays of $t - \delta t / 2$ and $t + \delta t / 2$. One can then vary the delay $t$ over the range where there is a signal. Subsequent integration gives the full relaxation curve at a given temperature. Typical results  are shown in Fig.\ \ref{fig_TcNqpZPD}(c). Repeating the measurement at various temperatures allows us to determine the thermal evolution of the  signal amplitude and effective relaxation time. These two quantities, however, can also be determined measuring only the maximum (positive) and minimum (negative) points in the derivative signal. These points are  indicated in Fig.\ \ref{fig_TcNqpZPD}(c) by `A' and `B', respectively. In a linear approximation, the amplitude of the integrated decay curve [squares in Fig.\ \ref{fig_TcNqpZPD}(c)] is proportional to the maximum slope (point `A'). This amplitude is related to the number of excess QPs ($N_{qp}$) through curves like the one in Fig.\ \ref{fig_TcNqpZPD}(b), yielding
\begin{equation}
  N_{qp}(T) \propto A \times \frac{\lambda^2(0)}{\lambda^2(T)}
    \times \frac{\hand{T}_S(0)}{ \hand{T}_S(T) - \hand{T}_N}.
  \label{eq_NqpDiff}
\end{equation}
Using the synchrotron pulse width, $t_{VUV}$, (which determines the time resolution) and assuming that the decay is represented by a single exponential (at least in the beginning of the relaxation), the effective relaxation time is given by 
\begin{equation}
  \tau_{\it eff} = - t_{VUV} \times \frac{A}{B}.
  \label{eq_taueffDiff}
\end{equation}

The photoinduced spectra were also determined in a differential mode. As the lamellar spectrometer works in a step-scan mode, the dithering method described in the previous paragraph can be used straightforwardly. To take advantage of the fast scan mode in the Bruker interferometer we could not utilize a lock-in based detection system. To obtain the differential spectra we recorded a set of 33 spectra alternating the two laser delays ($t$ and $t + \delta t$). Each spectrum has $n$ scans except for the first and the last spectra which are taken at the same delay with $n/2$ scans. All the measurements at the delay $t$ are summed up in \hand{T} and all those taken at $t + \delta t$ are added into $\hand{T}^\prime$. We can then determine the differential spectrum as $-\Delta \hand{T} / \hand{T} = \left( \hand{T} - \hand{T}^\prime \right) / \hand{T}$. This procedure minimizes the influence of the synchrotron light intensity variations arising from storage ring current decay.

As final remark, we are interested in the low-fluence regime. To verify that we are in this limit we set the pump-to-probe delay so that the differential signal is maximum [point `A' in Fig.\ \ref{fig_TcNqpZPD}(c)]. A necessary but not sufficient condition for a system to be in the low-fluence regime, is to have this signal varying linearly with the number of incident photons. By changing the laser power we can determine the boundary of the linear regime. This procedure was done for each sample at the lowest temperature measured, where the number of thermally produced QPs is the lowest.

\section{THE EQUILIBRIUM FAR-INFRARED RESPONSE}
\label{sec_eqfir}

The transmission of light thorough a thin film and into a substrate of refractive index $n$ is given in the long-wavelength limit by 
\begin{equation}
\hand{T} = \frac{4 n}{\left( y_1 + n + 1 \right)^2 + y_2^2}.
\label{eq_T_film}
\end{equation}
$y = y_1 + i y_2 = Z_0  \sigma d$ is the film's dimensionless complex admittance with $\sigma$ being the optical conductivity, $d$ the film thickness and $Z_0 \approx 377~\Omega$ the vacuum impedance.\cite{Gao1991} Equation \ref{eq_T_film} is valid when the film thickness is much less than the penetration depth and the measurement wavelenght. 

The normal-state frequency-dependent infrared response of metals and alloys is fairly well described by the  Drude model. The electrodynamics of BCS superconductors was calculated by Mattis and Bardeen\cite{Mattis1958} for dirty limit superconductors ($1 / \tau \gg 2 \Delta$) and an extension of those calculations was proposed by Zimmermann \etal\cite{Zimmermann1991} for an arbitrary scattering rate. All these expressions, valid in the BCS weak coupling limit, give the ratio $\left( \sigma^S_1 + i \sigma^S_2\right) / \sigma_0$, where $\sigma^S$ is the superconducting state complex conductivity and $\sigma_0$ is the normal-state dc conductivity. The theory for the strong-coupling superconducting-state optical conductivity was produced by Nam.\cite{Nam1967} Ginsberg \etal\cite{Ginsberg1976} pointed out that the low frequency approximation for these expressions corresponds to a decrease in the condensate weight. This decrease manifests itself as the subtraction of a factor $2~\delta S / \pi \omega$ from the imaginary part of the Mattis-Bardeen conductivity.

Note that Eq.\ \ref{eq_T_film} depends only on the product $\sigma d$ and not on the individual values of these two quantities. From the normal state transmission one can determine straightforwardly the sheet resistance $R_\Box = 1 / \sigma_0 d$ which can, in turn, be used to calculate the absolute values of $\sigma^S d$. This procedure works rather well for our samples because they are in the dirty limit, making $\omega \ll 1/\tau$ over the entire far infrared, so that the normal-state optical conductivity is essentially identical to the  dc conductivity.

We show in Fig.\ \ref{fig_TsTn} the ratio $\hand{T}_S / \hand{T}_N$ for all our samples. (Subscripts or superscripts $S$ and $N$ correspond respectively to the superconducting and normal states.) The parameters used in the fits are given in Table \ref{tb_TsTn}. In each sample we fitted the lowest temperature curve by varying the gap and the strong coupling coefficient $\delta S$. The experimentally determined temperature was then used to obtain the Mattis-Bardeen behavior (without changing any other parameter) for the remaining sample data. Within experimental errors, the fitted transmissions agree with the data. The fitting parameters are also in agreement with known values for these materials. Hence, the Mattis-Bardeen formalism and also our approximation in Eq.\ \ref{eq_T_film} describe correctly the superconducting response of our films justifying their use in the analysis of the photoinduced state (Sec.\ \ref{subsec_PI}) and in evaluating the sample temperature (see the Appendix).

\section{THE EXCESS QUASIPARTICLE STATE}
\label{sec_excessQP}

\subsection{Theory}
\label{subsec_theory}

\begin{figure}
  \begin{center}
    \includegraphics[width=8cm]{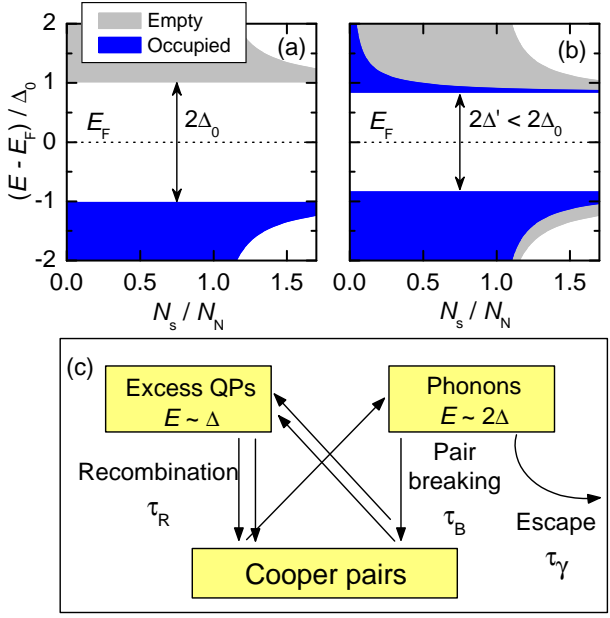} 
  \end{center}
\caption{(Color online) Ratio of the superconducting ($N_S$) and normal ($N_N$) densities of states at 0~K for a BCS superconductor (a) in equilibrium and (b) after photoexcitation and initial cascading effects but before QP recombination. The broken pairs in the photoexcited state produce the empty states below $E_F$ and occupied states above. Consequently, the superconducting gap is depleted. Panel (c) illustrates the relaxation process with competition between QPs and phonons.}
\label{fig_dynamics}
\end{figure}
Figure \ref{fig_dynamics} (a) shows the density of states at 0~K of a BCS superconductor. Light having a photon energy $h \nu > 2 \Delta_0$ is absorbed and breaks a Cooper pair, producing two QPs above the Fermi energy. In our experiments the photon energy is much higher than the superconducting gap and thus the QPs are created far above the Fermi level with an energy $\sim h\nu / 2$. Very quickly (in sub-picosecond times) these high-energy QPs relax {\it via\/} electron-electron and, eventually, electron-phonon scattering. These 
relaxation processes break more pairs, and the QPs quickly settle to energies close to $\Delta$. In such a cascading process, each QP initially created by the laser pulse will generate $\sim h \nu / 2 \Delta$ QPs close to the Fermi level. 

A similar branching process occurs in the phonon sector. Phonons of energy higher than $2 \Delta$ can break pairs, relaxing to $2 \Delta$ whereas phonons of energy less than $2 \Delta$  do not participate in the pair breaking process. Therefore, the system eventually reaches the state depicted in Fig.\ \ref{fig_dynamics}(b). This state is characterized by an excess (with respect to thermodynamic equilibrium) of unpaired QPs of energy $\Delta$  and, as shown by Owen and Scalapino,\cite{Owen1972} by a reduced energy gap.

The effective relaxation process must take into account a phonon bottleneck  effect, shown schematically in Fig.\ \ref{fig_dynamics}(c). Two QPs having energy $\Delta$ take a time $\tau_R$ to recombine into a Cooper pair. At recombination, a $2 \Delta$ phonon is emitted to carry the excitation energy of the QPs. The phonons with energy $2 \Delta$ can break a Cooper pair,  in characteristic time $\tau_B$, creating two QPs at the gap edge. A steady-state dynamic balance is established between the phonons and QPs. Eventually, however, the $2 \Delta$ phonons become depleted; either they relax anharmonically to lower energies or they leave the film. (The phonons can escape into the substrate or into their surrounding environment, such as the helium bath.) The phonon escape time is denoted $\tau_\gamma$.

The general case for any density of excess QPs has been discussed by Rothwarf and Taylor.\cite{Rothwarf1967} We are interested in the low-fluence limit, where the number of QPs created by the laser light is much smaller than the thermally broken pairs. In this case
the differential equations can be linearized giving:\cite{Gray1971}
\begin{equation}
  \frac{dN_{qp}}{dt} =  
    -\frac{2 N_{qp}}{\tau_R} + \frac{2 N_\Omega}{\tau_B}  
  \label{eq_dNqp}
\end{equation}
and
\begin{equation}
  \frac{dN_\Omega}{dt} =  
    -\left(\frac{N_\Omega}{\tau_B} + \frac{N_\Omega}{\tau_\gamma}\right) 
    + \frac{N_{qp}}{\tau_R}.
  \label{eq_dNOmega}
\end{equation}
$N_{qp}$ is the number of excess QPs at energy $\Delta$ and $N_\Omega$ is the number of phonons with energy $2 \Delta$. Assuming exponential decays for $N_{qp}$ and $N_\Omega$, one can solve the system given by Eqs.\ \ref{eq_dNqp} and \ref{eq_dNOmega}. The effective relaxation ($\tau_{\it eff}$) time is given by:
\begin{eqnarray}
  \frac{1}{\tau_{\it eff}} & = & \frac{2 \tau_R^{-1} + \tau_B^{-1} + \tau_\gamma^{-1}}{2} \times \nonumber \\
 		& & \times \left[ 1 - \sqrt{1 - \frac{8 \tau_\gamma^{-1} \tau_R^{-1}}{(2 \tau_R^{-1} + \tau_B^{-1} + \tau_\gamma^{-1})^2}}\,\right].
  \label{eq_taueff}
\end{eqnarray}
%

\begin{figure}
  \begin{center}
    \includegraphics[width=8cm]{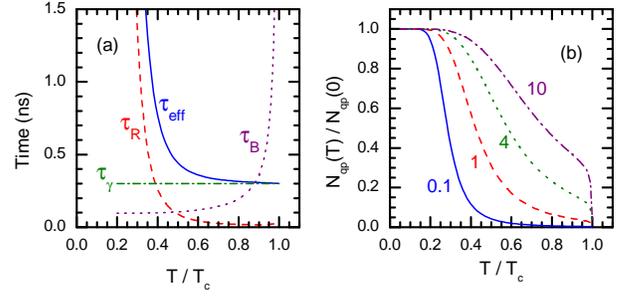} 
  \end{center}
\caption{(Color online) (a) Effective relaxation time ($\tau_{\it eff}$) calculated from the intrinsic QP recombination time ($\tau_R$), the pair breaking time by phonons ($\tau_B$) and the phonon escape time ($\tau_\gamma$) using Eqs.\ \ref{eq_taueff}, \ref{eq_tauR} and \ref{eq_tauB} with $\tau_R^0 = \tau_B^0 = 100$ ps. $\tau_\gamma = 300$ ps is taken to be temperature independent. (b) Temperature dependence of the excess number of QPs from Eq.\ \ref{eq_NqpT}. The dashed line was calculated using the same parameters as those used in panel (a). For the other curves, the number next to each line indicates the ratio $\tau_R^0 / \tau_B^0$ utilized.}
\label{fig_Kaplan}
\end{figure}
Another solution with the positive sign before the square root gives the time for phonons and QPs populations to reach equilibrium. Figure \ref{fig_Kaplan} (a) shows that for temperatures above $\sim 0.4 T_c$ but not very close to $T_c$ the phonon escape time $\tau_\gamma$ is much larger than $\tau_R$ or $\tau_B$. Hence Eq.\ \ref{eq_taueff} can be approximated by:
\begin{equation}
  \tau_{\it eff} = \tau_\gamma \left(1 + \frac{\tau_R}{2 \tau_B}\right).
  \label{eq_taueffapprox}
\end{equation}
In our numerical analysis we utilized the full expression given by Eq.\ \ref{eq_taueff}. However, the temperature range where Eq.\ \ref{eq_taueffapprox} is valid corresponds to the region covered by most of our experiments and we will use this expression for some qualitative analysis. 

\begin{figure*}
  \begin{center}
    \includegraphics[width=16cm]{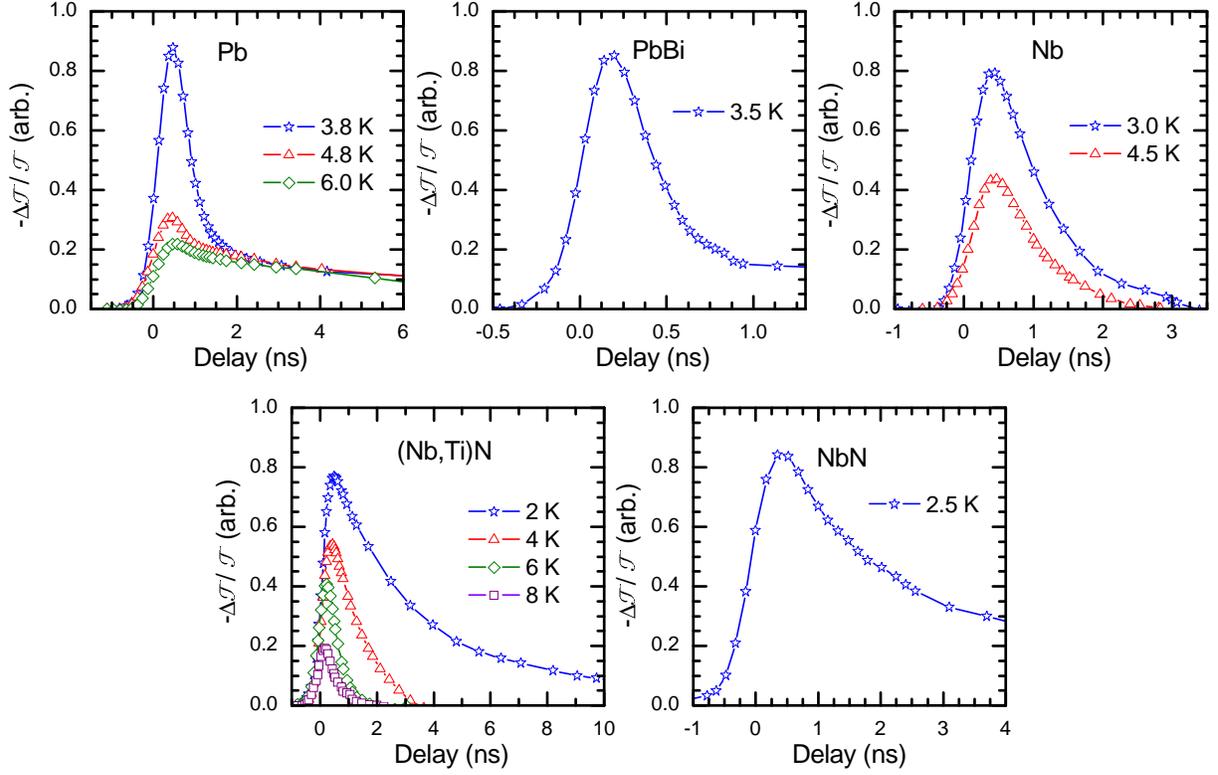} 
  \end{center}
\caption{(Color online) Relative variation in the overall far-infrared transmission $-\Delta\hand{T} / \hand{T}$ as a function of the delay between pump and probe pulses.}
\label{fig_Decay}
\end{figure*}

The phonon escape time depends on several factors such as the sample quality, the film thickness, acoustic mismatch between film and substrate, and the presence or absence of liquid helium surrounding the film.\cite{Kaplan1979} There is no detailed calculation for the temperature dependence of $\tau_\gamma$. However, at temperatures much smaller than the Debye temperature, the Kapitza resistance in a material interface behaves as $T^3$.\cite{NATO} As a higher Kaptiza resistance makes it harder for phonons to leave the material, in a first approximation we can assume that $\tau_\gamma = \tau_\gamma^0 + a T^3$. Several scattering rates for metallic superconductors were calculated by Kaplan \etal\cite{Kaplan1976} In particular, $\tau_R$ and $\tau_B$ can be obtained from intrinsic materials parameters:
\begin{widetext}
 \begin{equation}
  \frac{1}{\tau_R(\omega)} =  
    \frac{1}{1+f(\omega)}\left(\frac{\Delta_0}{k_BT_c}\right)^3 \frac{1}{\tau_R^0}
    \int_{\omega+\delta}^{\infty} d\Omega \quad \Omega^2 
    \frac{\Omega-\omega}{\sqrt{(\Omega-\omega)^2-\delta^2}} 
    \left[ 1 + \frac{\delta^2}{\omega(\Omega - \omega)} \right]
    \left[ n(\Omega)+1 \right] f(\Omega-\omega)    
  \label{eq_tauR}
\end{equation}
and
 \begin{equation}
  \frac{1}{\tau_B(\omega)} =  
    \frac{1}{\tau_B^0} 
    \int_\delta^{\omega-\delta} \frac{d\Omega}{\sqrt{\Omega^2-\delta^2}}
    \frac{\Omega(\omega-\Omega)+\delta^2}{\sqrt{(\omega-\Omega)^2-\delta^2}}
    \left[1 - f(\Omega) -f(\omega - \Omega)\right].
  \label{eq_tauB}
\end{equation}
\end{widetext}
Here, $\tau_R^0 = \hbar Z_1(0) / 2 \pi b (k_b T_c)^3$ and $\tau_B^0 = \hbar N / 4 \pi^2 N(0) \langle \alpha^2 \rangle \Delta_0$ are respectively the intrinsic pair recombination time close to $T_c/ 2$ and the phonon pair-breaking time at 0~K. All energies in Eqs.\ \ref{eq_tauR} and \ref{eq_tauB} are measured in units of $\Delta_0$. $\delta = \Delta / \Delta_0$ is the reduced gap,  $Z_1(0)$ is the QP renormalization factor, $N(0)$ is the electron single particle DOS at $E_F$, $N$ is the density of ions, and $\langle\alpha\rangle^2$ is the electron-phonon coupling function averaged over the whole phonon spectrum. In Eq.\ \ref{eq_tauR} the Eliashberg function $\alpha^2F(\Omega)$ is approximated by $b \Omega^2$, but the correct function can be used if it is known. $\omega$ is the energy of QPs just before recombination or the energy of photons available for pair breaking. $n(\Omega)$ and $f(\Omega)$ are the Bose and Fermi factors, respectively.

Figure \ref{fig_Kaplan}(a) shows $\tau_R$ and $\tau_B$ calculated from Eqs.\ \ref{eq_tauR} and \ref{eq_tauB}, assuming a weak coupling BCS temperature dependence for the gap, QPs with energy $\Delta$ and phonons with energy $2.1 \Delta$. We note that $\tau_R$ diverges at 0~K while $\tau_B$ remains finite. Therefore, at low temperatures, all the photon energy absorbed in the pair breaking process will appear as excess QPs. At higher temperatures the photon energies will end up distributed between QPs and phonons so that
\begin{equation}
  N_{qp}(0) \Delta_0 = N_{qp}(T) \Delta(T) + N_\Omega(T) 2\Delta(T).
  \label{eq_energybalance}
\end{equation}
When the phonon and QPs populations are in equilibrium, $N_{qp}/\tau_R \approx N_\Omega / \tau_B$ and, thus,
\begin{equation}
  \frac{N_{qp}(T)}{N_{qp}(0)} = 
    \frac{\Delta_0}{\Delta(T)} \frac{1}{1 + 2 \tau_B / \tau_R}.
  \label{eq_NqpT}
\end{equation}
With Eqs.\ \ref{eq_taueff} (or \ref{eq_taueffapprox}) and \ref{eq_NqpT} we can describe both the excess number of QPs and the effective relaxation time as a function of the temperature. Simulations of the effective relaxation time and number of excess QPs are shown in Fig.\ \ref{fig_Kaplan}. Note that whereas the effective relaxation time depends on the phonon escape time (a parameter that varies from sample to sample), the excess number of QPs depends only on the ratio $\tau_R^0 / \tau_B^0$. In addition, its strong dependence on $\tau_R^0 / \tau_B^0$ allows for an accurate determination of this ratio.

\subsection{Integrated response}
\label{subsec_integresp}

Figure \ref{fig_Decay} shows the overall photoinduced far-infrared  transmission for five films as a function of the time between pair breaking laser pulse and far-infrared probe pulse. For the samples where the data have been taken on a cold finger cryostat (Pb and Pb$_{0.75}$Bi$_{0.25}$) there is an initial fast relaxation followed by a long-lived tail. Data taken with the samples in contact with liquid or gaseous He [Nb, Nb$_{0.5}$Ti$_{0.5}$N and NbN] do not show this long-lived tail. However, this tail reappears in the data for the Nb sample when measured on the cold-finger cryostat. We therefore assign this long relaxation time process to film and substrate thermal effects rather than to intrinsic properties of the superconducting films. Heating issues are very important in pump-probe measurements and are discussed thoroughly for our samples in the Appendix.

Disregarding the long-lived tail in the Pb and Pb$_{0.75}$Bi$_{0.25}$ samples, the relaxation back to equilibrium can be described in all samples and at any temperature by a single exponential. We can then use the approximations given by Eqs.\ \ref{eq_NqpDiff} and \ref{eq_taueffDiff} to study the temperature dependence of the excess number of QPs and the effective relaxation rate.

\begin{figure}
  \begin{center}
    \includegraphics[width=8cm]{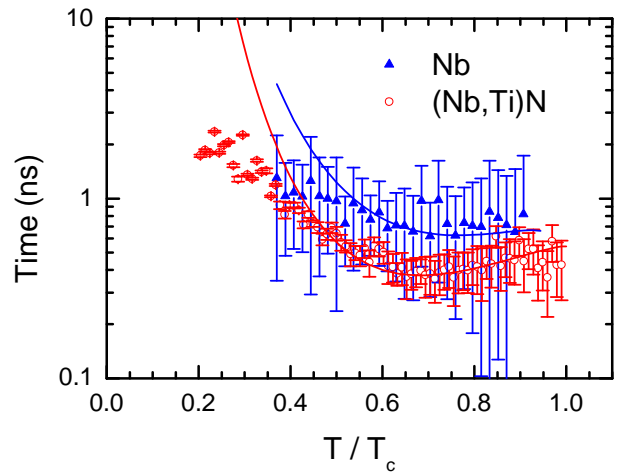} 
  \end{center}
\caption{(Color online) Effective relaxation time for Nb and Nb$_{0.5}$Ti$_{0.5}$N. The points are the experimental data transformed using Eq.\ \ref{eq_taueffDiff} and the solid lines are simulations using Eq.\ \ref{eq_taueff}. In both cases a small $T^3$ contribution to $\tau_\gamma$ is necessary to describe the rising part of the data above $0.8~T / T_c$. The error bars are obtained from statistical standard deviations from approximatelly 500 measurements at each temperature.}
\label{fig_taueff}
\end{figure}
Figure \ref{fig_taueff} shows the effective relaxation time as a function of temperature for the Nb and Nb$_{0.5}$Ti$_{0.5}$N samples. The fits describe the data properly at temperatures above $T_c / 2$ but  fail below this temperature, with the measured times being smaller that the
calculations. One possible explanation for the discrepancy relates to the QP energy before recombination. In our calculations we assumed that the QPs were sitting at an energy $\Delta$ prior to recombining into pairs and we disregarded recombination of higher energy QPs. Calculations using Eq.\ \ref{eq_tauR} show that $\tau_R$ decreases for increasing $\omega$. Indeed, it is possible to improve the fits between 0.2 and 0.6 $T_c$ in Fig.\ \ref{fig_taueff} by increasing $\omega$. This happens, however, at the expense of the fitting quality at higher temperatures. A complete description of this effective time may require recombination of QPs sitting at a temperature dependent distribution of energies. Another possibility is the effect of the smaller number of thermally broken pairs at low temperatures pushing the system into the high-fluence regime: the increased number of unpaired QPs available decreases the recombination time.

\begin{figure}
  \begin{center}
    \includegraphics[width=8cm]{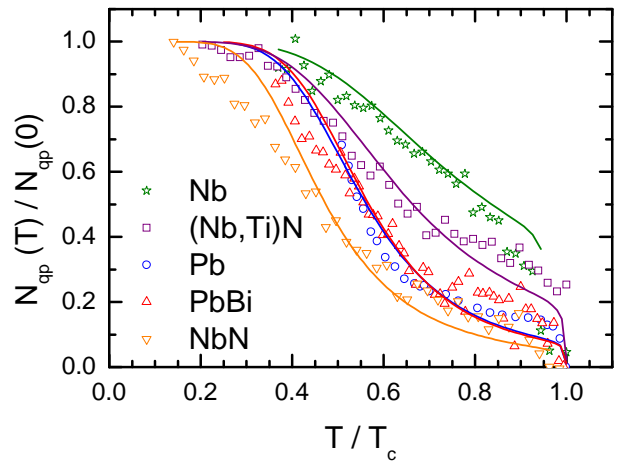} 
  \end{center}
\caption{(Color online) Thermal dependence of the number of excess QPs in all our samples. The solid lines are fits using Eq.\ \ref{eq_NqpT}.}
\label{fig_nqp}
\end{figure}
\begin{figure}
  \begin{center}
    \includegraphics[width=6cm]{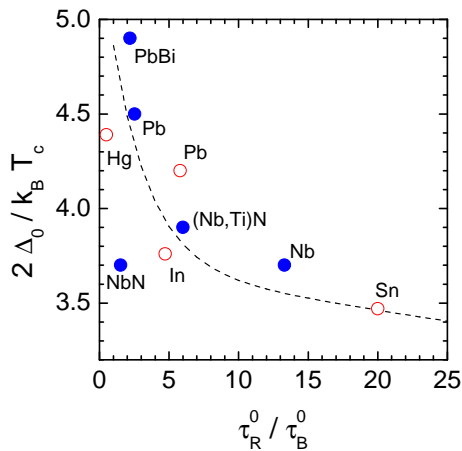} 
  \end{center}
\caption{(Color online) $2 \Delta_0 / k_B T_c$ plotted against $\tau_R / \tau_B$. The solid symbols are the results of this work. The open symbols are obtained from Kaplan \etal\protect\cite{Kaplan1976} calculations. The dashed line is a guide for the eye.}
\label{fig_strcpl}
\end{figure}

The data in Fig.\ \ref{fig_nqp} was calculated using Eq.\ \ref{eq_NqpDiff} and then corrected using curves similar to the one in Fig.\ \ref{fig_TcNqpZPD}(b). The fits for $\tau_{\it eff}(T)$ and $N_{qp}(T)$ do not allow for a determination of the absolute values of $\tau_R^0$ and$\tau_B^0$ as Eqs.\ \ref{eq_taueffapprox} and \ref{eq_NqpT} depend only on the ratio of these two time constants. This ratio however, is representative of the strong coupling character of the system. As shown in Fig.\ \ref{fig_strcpl}, a long phonon pair breaking time $\tau_B$ compared to $\tau_R$ corresponds to a large $2 \Delta_0 / k_B T_c$ ratio.

\subsection{Photoinduced spectra}
\label{subsec_PI}

The measurements discussed up to now were all integrated photoinduced transmission; no spectral features can be studied in this approach. Although those measurements determine the relation between the characteristic system relaxation times, one obviously cannot get direct information on the gap depletion predicted by Owen and Scalapino.\cite{Owen1972} By spectroscopically resolving the probe pulses we can compare the excited state and equilibrium spectra and directly measure the superconducting gap weakening.

In this case, we are mostly interested in the photoinduced response with pump and probe in coincidence. Although one can take measurements at several pump-to-probe delays, the time dependence of the gap depletion follows the effective relaxation time $\tau_{\it eff}$, already determined in the integrated measurements. Using the present synchronized method to measure the photoinduced spectra has the advantage of minimizing long-lived lattice heating effects and detecting the true changes produced by the excess QPs state.

Equation \ref{eq_T_film} shows that the superconducting-state transmission depends on the optical conductivity. A slight changes in $\sigma$ corresponds to changes in the transmission as:
\begin{equation}
  -\frac{\Delta \hand{T}}{\hand{T}} = \frac{\hand{T}(\sigma + \delta\sigma)
  - \hand{T}(\sigma)}{\hand{T}(\sigma)},
  \label{eq_dTT}
\end{equation}
where $\delta\sigma = \sigma(\Delta + \delta\Delta) - \sigma(\Delta)$. $\delta \Delta$ is the change in the gap arising from photoexcitation. In this picture, the equilibrium response Mattis-Bardeen equations can be used to describe a photoinduced state characterized by a gap weakening.

\begin{figure}
  \begin{center}
    \includegraphics[width=8cm]{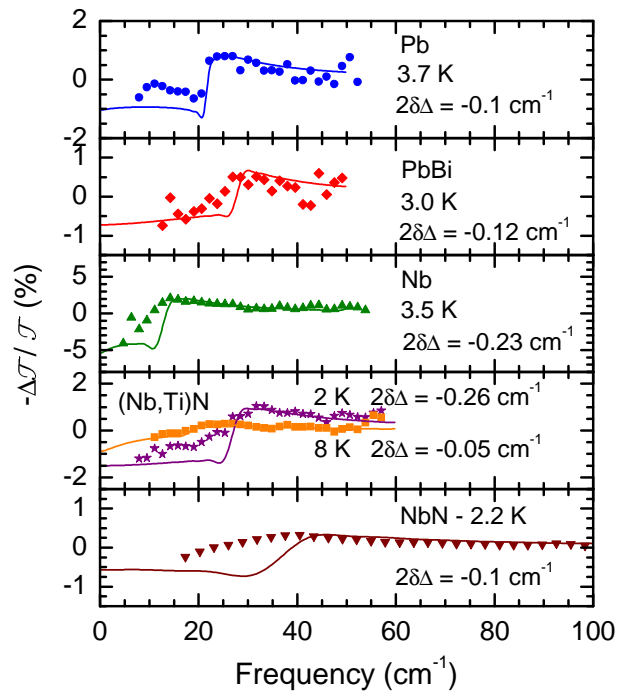}
  \end{center}
\caption{(Color online) Photoinduced spectra (pump and probe in coincidence) for all materials studied. The symbols are the experimental data and the lines are fits using Eq.\ \ref{eq_dTT} and Mattis-Bardeen optical conductivities.}
\label{fig_PIspectra}
\end{figure}
Figure \ref{fig_PIspectra} depicts a few $\Delta \hand{T} / \hand{T}$ spectra around the gap energy using the pump fluences shown in Table \ref{tb_TsTn}. The solid lines are simulations using Eq.\ \ref{eq_dTT} but keeping the same parameters used in the fits for $\hand{T}_S / \hand{T}_N$. These differential spectra allow us to determine directly the gap shift ($2~\delta \Delta$) produced by the excess QPs state. The magnitude of $2~\delta \Delta$ controls the height of the jump around $2 \Delta$ where $-\Delta \hand{T} / \hand{T}$ changes sign. The BCS fits are able to describe the data above $2 \Delta$ and the size of the $-\Delta \hand{T} / \hand{T}$ jump. Conversely, the quality of the low-energy fit is generally poor, with the data not showing as sharp a discontinuity as predicted by the theory. However, the basic physics appears to be what is expected: the superconducting gap is shifted to lower energies in the presence of the laser light, as the superfluid density is depleted and the zero-frequency superconducting delta function is weakened.

The knowledge of the gap shift also allows us to verify whether or not our data were in the low-fluence regime. In BCS theory, the superfluid concentration $n_s$ is of the order of $\Delta$ for $T \ll T_c$ and $\Delta^2$ for $T \approx T_c$. The ratio $\delta \Delta$ then tells us how many excess QPs we have in the system and this number can be compared to $1 - \lambda^2(0) / \lambda^2(T)$ representative of the number of thermally broken pairs $N_T$. The low-fluence limit is then given by the condition
\begin{equation}
  \frac{N_{qp}}{N_T} = \frac{\delta \Delta}{\Delta} \left[ 1 -
\frac{\lambda^2(0)}{\lambda^2(T)} \right]^{-1} \ll 1.
  \label{eq_lowfluencelim}
\end{equation}

Values for $N_{qp} / N_T$ are shown in Table \ref{tb_TsTn} at the lowest temperature measured in each sample. We notice that the low-fluence limit is indeed fulfilled for Pb, Pb$_{0.75}$Bi$_{0.25}$ and Nb at all temperatures but not for low temperatures in Nb$_{0.5}$Ti$_{0.5}$N and NbN. This effect may then explain the failure of our simple description of $\tau_{\it eff}$ at low temperatures and the better fits for  $-\Delta \hand{T} / \hand{T}$ obtained in Pb, Pb$_{0.75}$Bi$_{0.25}$ and Nb. It is also compatible with a poor $-\Delta \hand{T} / \hand{T}$ fit at low temperatures for Nb$_{0.5}$Ti$_{0.5}$N and a good fit for the same sample at 8~K. Indeed, at this temperature we calculated that $N_{qp} / N_T \approx 10^{-5}$. The understanding of the high fluence regime requires more data but deviations from our observations here are expected. In cuprate superconductors, for example, it was shown that the recombination dynamics is strongly dependent on the number of broken pairs.\cite{Gedik2004} 

\section{CONCLUSIONS}
\label{sec_conc}
We measured the equilibrium and photoinduced time dependent far-infrared response of several Pb, Pb$_{0.75}$Bi$_{0.25}$, Nb, Nb$_{0.5}$Ti$_{0.5}$N and NbN films in the superconducting state. We determined the effective relaxation time and number of excess quasiparticles created after the system has pairs broken by a fast laser pulse. Typical relations in the nanosecond range were observed. The most relevant quantity to characterize the relaxation process is the ratio between the intrinsic pair recombination time and the pair breaking time by phonons. The experimental results are in reasonable agreement with the linearized Rothwarf and Taylor\cite{Rothwarf1967} relaxation equations proposed by Gray.\cite{Gray1971} The gap shrinkage expected for the excess quasiparticle state was directly observed in the photoinduced far-infrared spectra. From a detailed analysis of the gap shift we can monitor the regions where the low-fluence limit approximation is valid.

\begin{acknowledgments}
This work was supported by the U.S. Department of Energy through contract DE-ACO2-98CH10886 at the NSLS. DOE also supported the operation of the synchronized laser system through contract DE-FG02-02ER45984 with the University of Florida. We are grateful to P.B. Allen, N. Bontemps, J. Carbotte, R. Combescot, T.P. Devereaux, M.V. Klein, I.I. Mazin, A.J. Millis, P. Monod, M.R. Norman, E. Nicol and M. Strongin for useful discussions.
\end{acknowledgments}

\appendix*
\section{Heating effects}
\label{app_heating}

One of the major experimental issues in all optical pump-probe spectroscopy is the ability to separate the intrinsic photoinduced effects from those produced by lattice heating. In our measurements, we observed two distinct lattice heating effects. The first is an average `static' heating produced by long term photon absorption. The second is a quick lattice heating that appears when the system is not able to dissipate the photon energy fast enough.

\begin{figure}
  \begin{center}
    \includegraphics[width=8cm]{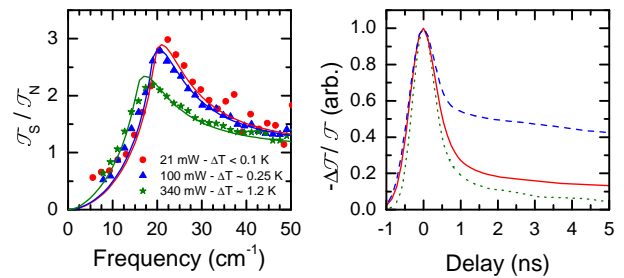} 
  \end{center}
\caption{(Color online) (a) $\hand{T}_S / \hand{T}_N$ for a Pb film as a function of laser power in CW mode measured at a nominal temperature of 4.77~K. The solid lines are BCS fits assuming higher effective lattice temperatures due to laser absorption. (b) Effective relaxation times for 3 different Pb films. The fast initial relaxation is the same for all films (the slight difference is due to different synchrotron pulse widths setting the time resolution). The long lived tail changes randomly from film to film and we assign it to different thermal contact conditions.}
\label{fig_heating}
\end{figure}

To evaluate and almost completely eliminate the `static' effect, we can use the changes in the far-infrared spectrum. This is done by setting the laser to its CW mode and measuring the far-infrared transmission as a function of laser power. In this case the laser pulses are at random times and therefore not synchronized to the infrared pulses. Alternatively one can also set the delay between pump and probe to a time much larger than the system excess QP relaxation time. In our samples both methods yielded the same results. Figure \ref{fig_heating}(a) shows a series of $\hand{T}_S / \hand{T}_N$ spectra for a Pb sample taken at identical temperatures but using different laser powers.
The increasing laser power changes the far-infrared transmission which can be described by a Mattis-Bardeen conductivity with a higher effective temperature. By fitting these spectra we can infer the temperature change produced by the laser and search for the laser power that minimizes this change. In our measurements, we made sure that this `average' lattice heating is less than 0.5~K. It is very important to remark that this effective lattice temperature is not the same $T^*$ discussed by Parker and Williams\cite{Parker1972}. In their formalism $T^*$ is an effective electronic temperature produced by an excess number of QPs. In our heating determination, the origin of the effective temperature is at the heat produced by the decay of electron hole-pairs created by the laser in the film. 
   
The second heating effect is harder to discriminate from intrinsic processes. While discussing the decays for Pb and Pb$_{0.75}$Bi$_{0.25}$ in Fig.\ \ref{fig_Decay} we assigned a long lived tail on the time dependent $-\Delta \hand{T} / \hand{T}$ to lattice heating. We have two major motivations to make this statement. As these films have been grown {\it in situ} at 80~K and needed to be kept cold at all times, we were constrained to measure them in a cold finger cryostat. The heat dissipation in this case only passes through the area in contact with the cryostat tip. As we are measuring the sample transmission, this thermal contact does not happen with the area under illumination. As shown in Fig.\ \ref{fig_heating}(b), different Pb samples show the same short decay but the long lived tail changes from sample to sample indicating an extrinsic process. Because the sample thermalization depends on the thermal contact between film and substrate and between substrate and cold finger, and because these factors were not always identical, this long lived tail is likely to be the signature of the lattice heating. The second motivation to make this assignment comes from measurements on a Nb film. The relaxation curves for this sample was measured in both cryostats; the long tail was only present for measurements where the sample was in vacuum in a cold-finger cryostat. When the sample is in He gas or liquid, the area under illumination is being cooled down by the helium, improving the heat dissipation.

In summary, we can easily eliminate the `average' static sample heating by carefully checking the appropriate laser power used as a pump. A second, time dependent heating effect, can be identified by comparing the decays in different samples. This time dependent heating effect has not been observed in measurements made with the sample in contact with He liquid or gas and indicate that it originates in a bad energy dissipation through the cryostat cold finger.



\begin{thebibliography}{00}


\bibitem{Burstein1961} E. Burstein, D.N. Langenberg, and B.N. Taylor, Phys. Rev. Lett. {\bf 6}, 92 (1961).

\bibitem{Ginsberg1962} D.M. Ginsberg, Phys. Rev. Lett. 8, 204 (1962).

\bibitem{Schrieffer1962} J.R. Schrieffer and D.M. Ginsberg, Phys. Rev. Lett. {\bf 8}, 207 (1962).

\bibitem{Rothwarf1963} A. Rothwarf and M. Cohen, Phys. Rev. {\bf 130}, 1401 (1963).

\bibitem{Rothwarf1967} A. Rothwarf and B.N. Taylor, Phys. Rev. Lett. {\bf 19}, 27 (1967).

\bibitem{Gray1969} K.E. Gray, Philos. Mag. {\bf 20}, 267 (1969).

\bibitem{Miller1967} B.I. Miller and A.H. Dayem, Phys. Rev. Lett. {\bf 18}, 1000 (1967).

\bibitem{Levine1968} J.L. Levine and S.Y. Hsieh, Phys. Rev. Lett. {\bf 20}, 994 (1968).

\bibitem{Gray1969b} K.E. Gray, A.R. Long, and C.J. Adkins, Philos. Mag. {\bf 20}, 273 (1969).

\bibitem{Gray1971} K.E. Gray, J. Phys. F: Metal Phys. {\bf 1}, 290 (1971).

\bibitem{Tinkham1972} M. Tinkham, Phys. Rev. B {\bf 6}, 1747 (1972).

\bibitem{Testardi1971} L.R. Testardi, Phys. Rev. B {\bf 4}, 2189 (1971).

\bibitem{Parker1972} W.H. Parker and W.D. Williams, Phys. Rev. Lett. {\bf 29}, 924 (1972).

\bibitem{Owen1972} C.S. Owen and D.J. Scalapino, Phys. Rev. B {\bf 28}, 1559 (1972).

\bibitem{Sai1974} G.A. Sai-Halasz, C.C. Chi, A. Denenstein, and D.N. Langenberg, Phys. Rev. Lett. {\bf 33}, 215 (1974).

\bibitem{Hu1974} P. Hu, R.C. Dynes, and V. Narayanamurti, Phys. Rev. B {\bf 10}, 2786 (1974).

\bibitem{Parker1975} W.H. Parker, Phys. Rev. B {\bf 12}, 3667 (1975).

\bibitem{Johnson1991} M. Johnson, Phys. Rev. Lett. {\bf 67}, 374 (1991).

\bibitem{Federici1992} J.F. Federici, B.I. Greene, P.N. Saeta, D.R. Dykaar, F. Sharifi, and R.C. Dynes, Phys. Rev. B {\bf 46}, 11153 (1992).

\bibitem{Han1990} S.G. Han, Z.V. Vardeny, K.S. Wang, O.G. Symko, and G. Koren, Phys. Rev. Lett. {\bf 65}, 2708 (1990).

\bibitem{Reitze1992} D.H. Reitze, A.M. Weiner, A. Inam, and S. Etemad, Phys. Rev. B {\bf 46}, 14309 (1992).

\bibitem{Feenstra1997} B.J. Feenstra, J. Schützmann, D. van der Marel, R. Pérez Pinaya, and M. Decroux, Phys. Rev. Lett. {\bf 79}, 4890 (1997).

\bibitem{Kabanov1999} V.V. Kabanov, J. Demsar, B. Podobnik, and D. Mihailovic, Phys. Rev. B {\bf 59}, 1497 (1999).

\bibitem{Averitt2001} R.D. Averitt, A.I. Lobad, C. Kwon, S.A. Trugman, V.K. Thorsm{\o}lle, and A.J. Taylor, Phys. Rev. Lett. {\bf 87}, 017401 (2001).

\bibitem{Kaindl2002} R.A. Kaindl, M.A. Carnahan, J. Orenstein, D.S. Chemla, H.M. Christen, H.Y. Zhai, M. Paranthaman, and D.H. Lowndes, Phys. Rev. Lett. {\bf 88}, 027003 (2002).

\bibitem{Segre2002} G.P. Segre, N. Gedik, J. Orenstein, D.A. Bonn, R. Liang, and W.N. Hardy, Phys. Rev. Lett. {\bf 88}, 137001 (2002).

\bibitem{Gedik2004} N. Gedik, P. Blake, R.C. Spitzer, J. Oresntein, R. Liang, D.A. Bonn, and W.N. Hardy, Phys. Rev. B {\bf 70}, 014504 (2004).

\bibitem{Howell2004} P.C. Howell, A. Rosch, and P.J. Hirschfeld, Phys. Rev. Lett. {\bf 92}, 037003 (2004).

\bibitem{Carbotte2004} J.P. Carbotte and E. Schachinger, Phys. Rev. B {\bf 70}, 014517 (2004).

\bibitem{Xu2003} Y. Xu, M. Khafizov, L. Satrapinsky, P. K\'u, A. Plecenik, and R. Sobolewski, Phys. Rev. Lett. {\bf 91}, 197004 (2003).

\bibitem{Demsar2003} J. Demsar, R.D. Averitt, A.J. Taylor, V.V. Kabanov, W.N. Kang, H.J. Kim, E.M. Choi, and S.I. Lee, Phys. Rev. Lett. {\bf 91}, 267002 (2003).

\bibitem{Lobo2004} R.P.S.M. Lobo, J.J. Tu, E.M. Choi, H.J. Kim, W.N. Kang, S.I. Lee, and G.L. Carr, condmat/0404708.

\bibitem{Carr2000} G.L. Carr, R.P.S.M. Lobo, J.D. LaVeigne, D.H. Reitze, and D.B. Tanner, Phys. Rev. Lett. {\bf 85} 3001 (2000).


\bibitem{Barber1993} Z.H. Barber, M.G. Blamire, R.E. Somekh, and J.E. Evetts, IEEE Trans. Appl. Supercond. {\bf 3}, 2054 (1993).

\bibitem{Somal1996} H.S. Somal, B.J. Feenstra, J. Schutzmann, J.H. Kim, Z.H. Barber, V.H.M. Duijn, N.T. Hien, A.A. Menovsky, M. Palumbo, and D. van der Marel, Phys. Rev. Lett. {\bf 76}, 1525 (1996).

\bibitem{Lobo1999} R.P.S.M. Lobo, J.D. LaVeigne, D.H. Reitze, D.B. Tanner, and G.L. Carr, Rev. Sci. Instrum. {\bf 70}, 2899 (1999).

\bibitem{Henry1979} R.L. Henry and D.B. Tanner, Infrared Phys. {\bf 19}, 163 (1979).

\bibitem{Lobo2002} R.P.S.M. Lobo, J.D. LaVeigne, D.H. Reitze, D.B. Tanner, and G.L. Carr; Rev. Sci. Instrum. {\bf 73}, 1 (2002).


\bibitem{Gao1991} F. Gao, G.L. Carr, C.D. Porter, D.B. Tanner, S. Etemad, T. Venkatesan, A. Inam, B. Dutta, X.D. Wu, G.P. Williams, and C.J. Hirschmugl, Phys. Rev. B {\bf 43}, 10383 (1991).

\bibitem{Mattis1958} D.C. Mattis and J. Bardeen, Phys. Rev. {\bf 111}, 412 (1958).

\bibitem{Zimmermann1991} W. Zimmermann, E.H. Brandt, M. Bauer, E. Seider, and L. Genzel, Physica C {\bf 183}, 99 (1991).

\bibitem{Nam1967} S.B. Nam, Phys. Rev. {\bf 156}, 470 (1967).

\bibitem{Ginsberg1976} D.M. Ginsberg, R.E. Harris, and R.C. Dynes, Phys. Rev. B {\bf 14}, 990 (1976).


\bibitem{Kaplan1979} S.B. Kaplan, J. Low Temp. Phys. {\bf 37}, 343 (1979).

\bibitem{NATO} {\it Nonequilibrium Superconductivity, Phonons, and Kapitza Boundaries}, Ed. K.E. Gray, NATO Advanced Study Institutes Series, Plenum Press, New York and London (1981).

\bibitem{Kaplan1976} S.B. Kaplan, C.C. Chi, D.N. Langenberg, J.J. Chang, S. Jafarey, and D.J. Scalapino, Phys. Rev. B {\bf 14}, 4854 (1976).


\end{thebibliography}
\end{document}